\documentclass[sigconf]{acmart}

\AtBeginDocument{%
  }

\setcopyright{acmlicensed}
\copyrightyear{2026}
\acmYear{2026}
\acmDOI{XXXXXXX.XXXXXXX}

\acmConference[TBD]{}{}{}

\usepackage{booktabs}
\usepackage{tikz}
\usetikzlibrary{arrows.meta, positioning, bending}

\begin{document}

\title{AI-Generated Email Drafts Shift Culturally Distinctive Communication Styles in Professional Email}

\author{Shintaro Sakai}
\affiliation{%
  \institution{Indiana University Bloomington}
  \city{Bloomington}
  \state{Indiana}
  \country{USA}
}
\email{shinsaka@iu.edu}

\author{Alice Gao}
\affiliation{%
  \institution{University of Washington}
  \city{Seattle}
  \state{Washington}
  \country{USA}
}
\email{atgao@cs.washington.edu}

\author{Yuichi Shoda}
\affiliation{%
  \institution{University of Washington}
  \city{Seattle}
  \state{Washington}
  \country{USA}
}
\email{yshoda@uw.edu}

\author{Katharina Reinecke}
\affiliation{%
  \institution{University of Washington}
  \city{Seattle}
  \state{Washington}
  \country{USA}
}
\email{reinecke@cs.washington.edu}

\begin{abstract}
AI assistants that support email composition may shift cultural communication norms, such as the directness typical of low-context cultures like the US versus the indirectness and contextual sensitivity central to high-context cultures like Japan. Yet it remains unknown to what extent people adopt and edit AI drafts inconsistent with their cultural communication norms. 
We address this through a preregistered within-subject experiment in which Japanese and American participants wrote workplace emails in their native language without AI, with a low-context AI, and with a high-context AI.
We found that Japanese participants wrote emails with significantly more high-context markers (politeness, apologies) than Americans. But AI drafts shifted participants' emails toward the draft's style, with larger shifts when the draft was culturally misaligned: Japanese drifted most under low-context drafts, Americans most under high-context drafts. These findings suggest AI drafts risk overwriting cultural communication norms unless they adapt to users’ communication styles.
\end{abstract}

\begin{teaserfigure}
  \includegraphics[width=\textwidth]{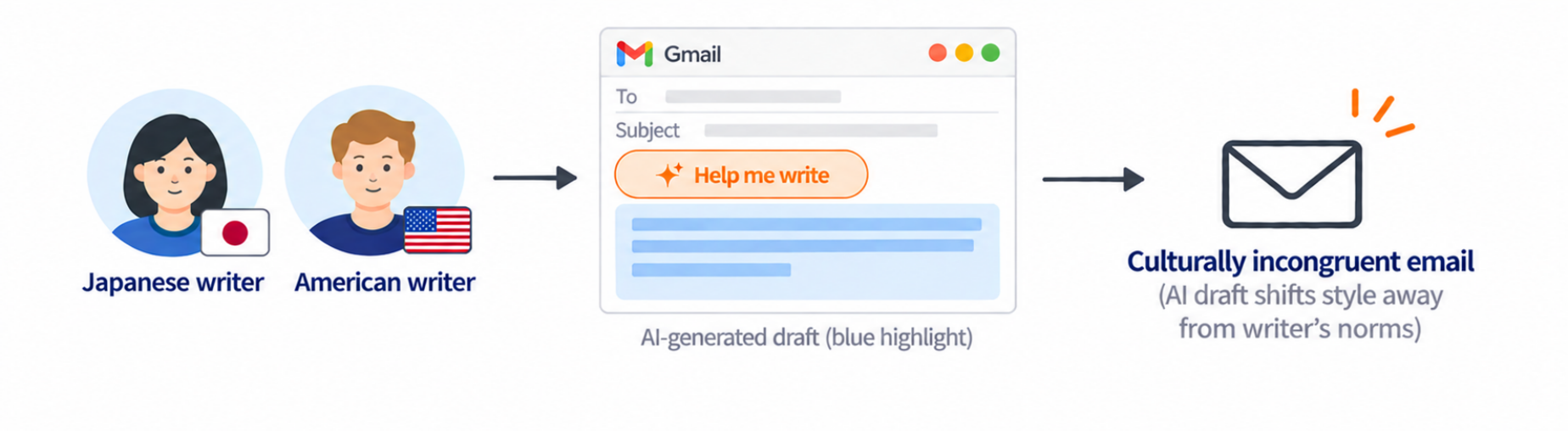}
  \caption{The results of an online experiment with Japanese and American participants showed that AI-assisted email drafts can produce culturally incongruent emails that shift a writer's style away from their cultural communication norms.}
  \label{fig:teaser}
\end{teaserfigure}



\maketitle


\section{Introduction}

Email is among the most common forms of professional communication, used daily to negotiate requests, deliberate about decisions, and maintain working relationships. Yet how people write emails is not merely a matter of personal preference: it is governed by culturally rooted communication norms that reflect deeper values about relationships, hierarchy, and meaning~\cite{holtbrugge2013cultural}. One influential framework for capturing these differences across cultures is Hall's distinction between high- and low-context communication~\cite{holtbrugge2013cultural, hall1976beyond}: in high-context cultures, such as Japan, meaning is conveyed indirectly and embedded in relational context; in low-context cultures, such as the United States, meaning is stated directly and explicitly. Adhering to these norms is essential for building trust and maintaining professional relationships. Conversely, violating them can create friction and damage working relationships, as documented in cross-cultural professional and team contexts~\cite{hinds2011global, tenzer2014impact, adair2024listening}.

AI writing assistants are now embedded in the tools professionals use to compose these emails. Gmail's ``Help me write'' feature can draft emails from scratch and adjust tone~\cite{google_helpmwrite}, and Microsoft Outlook integrates Copilot to draft emails from prompts, rewrite existing drafts, and adjust tone and length~\cite{microsoft_wordai}. As these tools become more widely adopted, a critical question emerges: could AI email assistants overwrite the culturally distinct communication norms that users and recipients rely on to build and maintain interpersonal relationships?

There is growing evidence that this might be the case. The most widely used LLMs have been found to exhibit systematic biases toward Western, and specifically American, cultural norms~\cite{cao2023assessing, tao2024cultural, alkhamissi2024investigating, rao2025normad}. In AI writing tools, such biases can shift users' word choice toward Western patterns and reduce vocabulary diversity~\cite{agarwal2025chi}. If large numbers of people rely on a small set of AI tools that share similar cultural values (exacerbated by an ``artificial hivemind'' effect~\cite{jiang2026hivemind}), this could progressively homogenize how people write. Researchers have described this influence on cultural norms as ``digital colonialism,'' highlighting how these tools can amplify power by those who exercise imperial control~\cite{couldry2019data, kwet2019digital}. Over time, these lopsided power dynamics could lead individuals and societies to adopt cultural narratives, norms, and values that are misaligned with their own, flattening how people write and communicate~\cite{agarwal2025chi}. 

Of course, AI models can only have this kind of influence if users are susceptible to adopting their suggestions, and with that, their biases. However, it remains unknown whether and to what extent people adopt AI-generated email drafts, and whether they would do so if the drafts are inconsistent with their cultural communication norms. This gap matters for two related reasons. First, prior work has documented AI-induced shifts in individual writing tasks~\cite{agarwal2025chi}, but workplace emails differ in ways that make this question especially consequential because they are directed at specific recipients and carry genuine social stakes: an email to a superior that violates cultural norms can damage working relationships or signal disrespect~\cite{hinds2011global, tenzer2014impact, adair2024listening}. Moreover, because emails are exchanged back and forth, AI-induced shifts in one person's email communication style may not only harm their own working relationships but also affect the expectations and communication norms of those they correspond with, cumulatively reshaping how people communicate~\cite{morling2016cultural}.

A second gap is the role of cultural alignment. It remains largely unknown whether AI's homogenizing influence is weaker or stronger depending on whether a user's cultural communication norms are more or less culturally distant from the AI's default. Much of the prior work has focused on a Global North/Global South divide, such as by examining how US models influence the output of Indian users~\cite{agarwal2025chi}. But the AI landscape is diversifying rapidly: many models now reflect non-Western values because of differing training data and annotator demographics~\cite{srivastava2026cost}. There are also growing efforts to develop LLMs tailored to specific local communities~\cite{el2025nilechat}. Understanding whether the degree of cultural misalignment between the AI and the user amplifies the shift becomes critical.

In the present study, we address both gaps experimentally. We ask: \textbf{(RQ1)} Does using an AI email draft shift the cultural communication style expressed in users' emails toward the style of the draft? \textbf{(RQ2)} And does the magnitude of this shift---and users' reliance on the AI draft---depend on the cultural alignment between the AI and the user's background? 

We answer these questions with a controlled, preregistered experiment involving 175 full-time Japanese and American employed adults (86 Japanese, 89 American). Each participant wrote three workplace emails: one without AI assistance, one with an AI offering a low-context draft, and one with an AI offering a high-context draft (counterbalanced). Participants' emails responded to one of three scenarios: a favor-request, a disagreement, and a task-refusal to a superior, situations known to elicit strongly differentiated communication norms across cultures~\cite{brown1987politeness, ting1982toward}.

Our results show that Japanese participants wrote significantly more high-context emails than Americans in the no-AI baseline condition, particularly by using imposition mitigation strategies such as more polite request forms and apologies. However, AI email drafts shifted the communication style of participants' emails toward the draft's context orientation: high-context AI drafts led participants to write more high-context emails, while low-context AI drafts shifted writing in the opposite direction. Importantly, this shift was larger when the draft was culturally misaligned with participants' cultural background---Japanese participants shifted much more under low-context drafts than high-context ones, and Americans shifted much more under high-context drafts than low-context ones. The finding suggests that culturally misaligned AI drafts are more disruptive than aligned ones, amplifying the shift in communication style. 

Overall, our results provide empirical evidence that AI writing assistants can change how emails are written and that their influence is strongest when a user's cultural communication style diverges from the AI's default. Critically, this risk is bidirectional --- it is not limited to Western AI shifting non-Western communication styles. 
Together, these findings point to a need for AI writing assistants that take into account users' cultural communication styles rather than defaulting to a single norm.

\paragraph{\textbf{Preregistration and Dataset}}
Our anonymized experiment preregistration can be accessed at: \url{https://aspredicted.org/2jy53q.pdf}.
In addition, our dataset of 771 workplace emails (474 American, 297 Japanese), including no-AI, low-context AI draft, and high-context AI draft conditions, is available for replication and future research at [omitted].

\section{Related Work}

\subsection{High- and Low-Context Communication Across Cultures}
\label{sec:highlow}

People's communication norms are rooted in the cultural values that members of each culture bring to a communicative
situation. In collectivistic cultures, more prevalent in East Asian countries, individuals
hold an interdependent self-construal: the self is defined in relation
to others, social harmony is prioritized, and conflict is viewed as a
threat to the broader social network
\cite{markus2014culture, hofstede2001cultures, morris1998missing}.
In individualistic cultures, more prevalent in Western countries,
individuals define themselves through unique internal characteristics and personal goals, and communication functions as a means of direct self-expression rather than relationship maintenance
\cite{markus2014culture}. These cultural values are reflected in a culture's dominant communication norms. A foundational framework for understanding this variation is Hall's high/low-context distinction \cite{hall1976beyond}. In high-context cultures, common in East Asian countries, meaning is conveyed implicitly and indirectly through shared context rather than expressed explicitly. In low-context cultures, more prevalent in Western countries, people favor explicit and direct expression, with meaning encoded fully in the words themselves.

High collectivism and high-context communication tend to travel together, because in tightly-knit, relational societies, group members share more implicit background knowledge. Communication can therefore rely on indirectness, inference, and reading between the lines without causing misunderstanding. Collectivist values also manifest in imposition mitigation: High-context communicators prioritize preserving the relationship, protecting the other person's face, avoiding overt imposition and open disagreement \cite{brett2007negotiating, trubisky1991influence, oetzel2003face, gudykunst1988culture, tinsley2004culture, adair2004culture, ting1991culture, ting1998facework, kim1998high}. They use hedging, pre-emptive apologies, and deference markers to soften the burden placed on the recipient \cite{brown1987politeness, lee2012cultural, lee2014effectiveness, gagne2010reexamining}. By contrast, individualistic communicators are more likely to state requests and disagreements directly, with minimal mitigation \cite{brett2007negotiating, kim1998high, lee2012cultural}.

These differences are especially salient in professional contexts: cultural differences in indirectness are amplified in work settings compared to social ones \cite{sanchez2003conversing}, and these patterns persist in professional email \cite{holtbrugge2013cultural}.

In the present study, we operationalize cultural communication style through two theoretically motivated components: indirectness and imposition mitigation. Although these components were developed through distinct theoretical traditions --- indirectness primarily through Hall's high/low-context communication theory~\cite{hall1976beyond}, and imposition mitigation through politeness theory~\cite{brown1987politeness} --- they are closely related in cross-cultural communication: high-context communication tends toward greater indirectness and stronger imposition mitigation, while low-context communication tends toward directness and less mitigation. In practice, the two components are highly overlapping and difficult to cleanly distinguish in annotation. We therefore treat them as jointly constituting a single composite measure of high-context communication style, rather than as independent dimensions.

\subsection{Cultural Bias and Its Consequences in AI Writing Assistance}

A growing body of research demonstrates that LLMs exhibit systematic bias toward Western, and specifically American, culture \cite{cao2023assessing, tao2024cultural, alkhamissi2024investigating, rao2025normad}. As LLM-powered AI tools become increasingly embedded in everyday life, these cultural biases may be reflected in their outputs and, in turn, influence the users who rely on them.

One domain where such influence may be particularly consequential is writing assistance. LLMs have been widely deployed to support writing across many domains \cite{liang2025widespread}. Prior research has shown that AI writing assistants shape users' output in consequential ways: LLM use produces large semantic shifts including increased argumentative neutrality \cite{abdulhai2026llms}, shifts users' expressed opinions toward the position advocated by the model \cite{williams2026biased}, and reduces the collective diversity of content produced across writers \cite{doshi2024generative, padmakumar2024does}. These findings raise concerns about how culturally biased AI systems may shape the way people write. Work by Agarwal et al.~\cite{agarwal2025chi} finds that AI autocomplete suggestions led Indian participants to adopt American writing patterns, shifting not only the cultural content of what they wrote, but also stylistic features such as lexical diversity. Their work showed that people tend to retain many of the AI's suggestions in a general writing task, but that Indian participants tended to accept the AI's suggestions at significantly higher rates (25\%) than Americans (19\%)~\cite{agarwal2025chi}, a differential effect that was also found by Gao et al.~\cite{gao2026framing}. 
We extend this prior work by exploring how people may engage with AI email drafts that, unlike autocomplete sentences, are presented as a whole, and how this interaction with AI could impact a person's cultural  communication norms that are essential to relationship maintenance, face management, and the negotiation of meaning across cultural contexts. 

\section{Method}

This study was pre-registered on AsPredicted prior to data collection and received Institutional Review Board approval (details omitted for anonymous review).

\subsection{Hypotheses}

We pre-registered four hypotheses: 

\noindent\textbf{H0:} Japanese participants will write emails in a more high-context style than American participants in the no-AI baseline condition. \\
\noindent\textbf{H1:} Participants who use an AI email draft will shift the communication norms of their emails in the direction of the draft's context orientation, relative to the no-AI baseline. \\
\noindent\textbf{H2:} This shift will be larger when the AI draft is culturally misaligned with the participant's background. That is, Japanese participants will show a larger shift under the low-context draft, and American participants will show a larger shift under the high-context draft. 

Beyond these stylistic shifts, we also examined whether cultural alignment shapes participants' behavioral engagement with the draft itself. Culturally aligned drafts may feel more natural and require fewer edits, leading to higher AI reliance. \\
\noindent\textbf{H3:} AI reliance will be higher when the draft is culturally aligned with the participant's background than when it is misaligned. Specifically, Japanese participants will retain more of a high-context draft than a low-context draft, and American participants will retain more of a low-context draft than a high-context draft.

\subsection{Design and Procedure}
\label{sec:procedure}

The study used a 2 (culture: Japanese vs.\ American) $\times$ 3
(condition: No-AI, AI Low-Context Draft, AI High-Context Draft) mixed design, with
culture as a between-subjects factor and condition as a within-subjects
factor. Japanese participants completed their tasks in Japanese; American participants in English.
To prevent carryover effects from writing the same email repeatedly, each
task used a different scenario type.
Scenario types were rotated across task positions using a Latin square
(Groups A, B, C), and the order of the two AI draft types (Low-Context vs.\
High-Context) across Tasks~2 and~3 was independently counterbalanced
(Conditions A and B), yielding six unique combinations for each culture (see Appendix~\ref{app:design}).

\begin{figure}[h]
  \centering
  \includegraphics[width=0.6\linewidth]{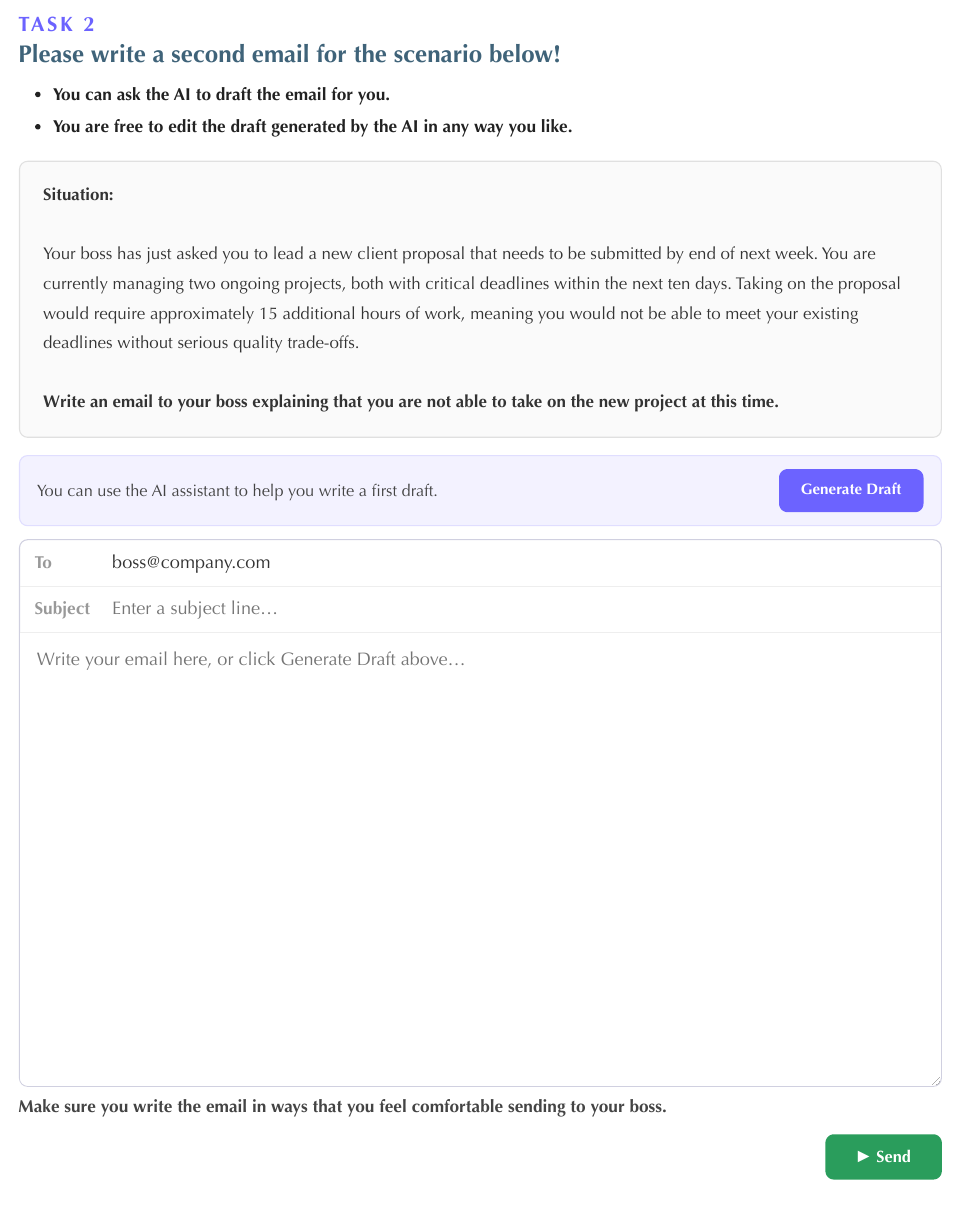}
  \caption{The email composition interface presented to participants in the AI conditions (Tasks~2 and~3). Participants could click ``Generate Draft'' to receive the pre-generated AI draft, then edit it freely or write from scratch before submitting.}
  \label{fig:interface}
\end{figure}

After providing informed consent,
participants were introduced to the supervisor framing (described in Section~\ref{sec:scenarios}) and then completed three email writing tasks,
each presenting the assigned scenario description in the upper half of the screen alongside the email composition interface below. In Task~1 (No-AI), they wrote the email entirely from scratch using a plain email composition interface.
In Tasks~2 and~3 (AI conditions, see Figure~\ref{fig:interface}), the same interface included a
``Generate Draft'' button that, when clicked, displayed the relevant
pre-generated AI draft. Participants could edit the draft freely before
submitting, or ignore it and write from scratch. To increase ecological validity, before submission, a confirmation dialog asked ``Are you sure you want to send this email to your boss?''. The final submitted email
was recorded and used for the analysis. 

After completing all three email tasks, participants answered a brief
survey including AI use frequency in professional
email writing, and standard demographic items (age, gender, education);
see Appendix~\ref{app:survey}.

\subsection{Email Writing Scenarios and AI Drafts}
\label{sec:scenarios}

The three workplace scenario types are \emph{favor-requests}, \emph{disagreements}, and \emph{refusals}. All scenarios were framed as writing to a supervisor with whom the participant had worked for several years and had a well-established relationship. These types and framings were selected because each involves a face-threatening act in a hierarchical relationship, making cultural differences in communication norms particularly likely to emerge \cite{brown1987politeness, ting1982toward}. This shared framing was chosen to hold the relational context constant across participants. The full scenarios and detailed scenario selection procedure are provided in Appendix~\ref{app:scenarios}.

Each AI condition presented a pre-generated high- or low-context email draft for the participant's assigned scenario.
These drafts were generated using three LLMs (GPT-5.5, Claude Sonnet 4.6, and Gemini 2.5 Flash) rather than crafted manually by the authors, ensuring higher ecological validity.
Every participant in the same condition and scenario saw the identical draft to preserve stimulus control. The draft and scenarios selection procedure intentionally prioritized stylistic contrast: for each scenario, the candidate scenario and draft pair yielding the largest difference between the high- and low-context versions were selected as study stimuli to give the manipulation the best opportunity to produce an observable effect on participants' writing. Table~\ref{tab:draft_examples} shows the two AI drafts used in the experiments for the favor-request scenario (English versions).
The detailed draft selection procedure and full English AI email drafts are provided in Appendix~\ref{app:scenarios} and~\ref{app:drafts}.

As reported in Table~\ref{tab:draft_scores} (Appendix~\ref{app:scenarios}), high-context scores (quantified using the measure described in Section~\ref{sec:measures}) varied across the three LLMs: GPT-5.5 produced the least high-context output in both English and Japanese, while Claude and Gemini scored higher in both languages. Prompt language also shaped context-orientation: drafts generated from Japanese prompts scored substantially more high-context than those from English prompts across all three models, consistent with Japanese being a high-context language.

The selected scenario and email drafts were back-translated by a Japanese--English bilingual speaker so that American participants received English versions and Japanese participants received Japanese versions. A separate native speaker of Japanese and a native speaker of English each reviewed the translated scenarios and drafts in their respective language, correcting any expressions that read unnaturally without changing the high- or low-context level of the email drafts.

\begin{table}[h]
\small
\caption{AI drafts for the favor-request scenario (English versions shown; Japanese participants received back-translated equivalents).}
\label{tab:draft_examples}
\begin{tabular}{p{0.46\linewidth}p{0.46\linewidth}}
\toprule
\textbf{Low-context draft} & \textbf{High-context draft} \\
\midrule
{\itshape Subject: Request for Leave This Week\par\smallskip
Hi [Supervisor's Name],\par\smallskip
I hope this email finds you well. I am writing to request three days of leave from Wednesday, November 8th, through Friday, November 10th.\par\smallskip
A very close friend is getting married this Saturday, and I was only recently asked to be a part of the wedding party. This requires me to travel out of town early to participate in necessary preparations.\par\smallskip
I understand this is incredibly short notice and goes against your usual policy for last-minute requests, which I fully respect. However, missing this significant event for a close friend is truly not an option for me.\par\smallskip
I am already proactively preparing for my absence by completing urgent tasks and ensuring comprehensive handovers for all critical responsibilities. I am committed to minimizing any disruption.\par\smallskip
Thank you for considering my urgent request.\par\smallskip
Sincerely,\newline
[Your Name]}
&
{\itshape Subject: Request for Leave This Week\par\smallskip
Hi [Supervisor's Name],\par\smallskip
I apologize for contacting you while you are away. I sincerely apologize for this sudden message, but I am writing to request three days of leave from this Thursday evening through Sunday.\par\smallskip
A long-time close friend's wedding is this Saturday, and last week I was suddenly asked to attend as a member of the wedding party. This has resulted in a very last-minute request, and I am deeply sorry.\par\smallskip
I am fully aware that this last-minute leave request goes against your policy, and I am truly sorry. However, I humbly ask for your permission so that I may fulfill this important role in celebrating my close friend's new chapter.\par\smallskip
To ensure there is no disruption to work during my absence, I will complete my projects as much as possible in advance, ask my colleague [Name] to handle urgent matters, and ensure a thorough handover.\par\smallskip
I apologize for this selfish request, but I sincerely hope you will approve it.\par\smallskip
Respectfully,\newline
[Your Name]} \\
\bottomrule
\end{tabular}
\end{table}

\subsection{Measures}
\label{sec:measures}

\subsubsection{Communication Style}

To quantify the level of high-context communication in each email, we counted the number of linguistic markers associated with the two components introduced in Section~\ref{sec:highlow} --- indirectness and imposition mitigation. Marker counts were normalized by total word count to yield a per-100-word score for each email. The two normalized scores were then summed to produce a single composite high-context score, with higher values indicating a more high-context communication style.

The linguistic markers were drawn from prior work on cross-cultural professional communication \cite{holtbrugge2013cultural, wang2010rome, hara2010cross, lee2012cultural}. Because no single prior work provides a comprehensive marker set covering both components, we supplemented the literature-derived list with additional candidate markers generated by Claude Sonnet 4.6. We prompted Claude with the theoretical definitions of both components, to generate additional candidate markers for human review. All candidate markers were reviewed by two authors with expertise in high/low-context communication theory. Each candidate was assessed against the theoretical definition of its component, and only markers on which both authors reached agreement were retained in the final set. The complete list of markers is provided in Appendix~\ref{app:markers}. To validate that the markers capture the intended construct, two authors blind to the scores were presented with the three pairs of AI drafts used in the experiment, one high-context and one low-context draft per scenario, and independently judged which draft in each pair was more high-context. Both authors' judgments agreed with the score rankings in all three cases.

Since Japanese participants wrote their emails in Japanese, these were translated into English prior to scoring so that a single coding scheme could be applied uniformly across both groups to preserve construct validity. To ensure translation quality, we followed the same back-translation verification procedure described in Appendix~\ref{app:scenarios}.

Markers were counted by a single bilingual coder, who cross-referenced the original Japanese transcripts when scoring translated emails. We also note that marker categories are hard to cleanly distinguish and some expressions can satisfy the criteria of more than one category simultaneously. In such cases, each expression was counted only once — assigned to the most fitting category to avoid double-counting and inflating the composite score. To verify coding consistency, Python-based automated checks were run across both American and Japanese emails to identify and correct any phrases scored or categorized inconsistently. Applying these checks uniformly across both groups ensures that the coding scheme was applied with equal consistency to American and Japanese emails, preserving the validity of the cross-cultural and within-person comparisons that form the basis of our results.

\subsubsection{AI Reliance}

To quantify how much participants relied on the AI draft, we computed an \emph{AI reliance score} for each submitted email.
Following previous work~\cite{agarwal2025chi}, we define AI reliance as the proportion of the final email that overlaps with the AI draft:
\[
  \text{AI Reliance} = \frac{\text{number of AI-draft words retained in the final email}}{\text{total words in the final email}}
\]
While the previous work~\cite{agarwal2025chi} computed this metric at the character level, which is well-suited to their inline-autocomplete setting, we compute it at the word level as participants received a pre-generated full draft and edited it freely in a text box. A score of~1 means the participant submitted the AI draft as-is, and a score of~0 means no AI text was retained.

\subsection{Participants}
\label{sec:participants}

We recruited 257 full-time employed adults through Prolific: 99 Japanese
participants in Japan and 158 American participants in the United States.
Because generating the AI draft was optional in our experiment design (see Section~\ref{sec:procedure}), some participants did not generate the draft in Task 2 or Task 3. To maintain a consistent analytic sample across all hypotheses, we only kept participants who generated the AI draft in both AI conditions: 87 Japanese and 90 American participants.
This sample size was determined by a priori power analysis targeting a medium effect size (Cohen's $f = 0.25$, $\alpha = .05$, power $= .80$); details are provided in Appendix~\ref{app:power}. Full-time employment was required
because the experimental scenarios are grounded in everyday workplace
situations. Participants who did not meet this criterion were screened
out at the eligibility stage.
As an attention check, participants whose submitted emails did not match the assigned scenario were excluded from analysis; one Japanese and one American participant were excluded on this basis.
American participants had a mean age of 38.0 years ($SD = 11.0$) and Japanese participants 41.3 years ($SD = 10.6$); the groups differed significantly in age, though the difference was small ($t(173) = -2.05$, $p = .042$). Gender distribution (American: 47 female, 41 male, 1 non-binary; Japanese: 30 female, 55 male, 1 non-binary) and education level did not differ significantly between groups ($\chi^2 = 5.75$, $p = .057$; $t(173) = 1.66$, $p = .098$, respectively). Further demographic details (education, company tenure, and AI writing tool use) are provided in Table~\ref{tab:demographics} in Appendix~\ref{app:demographics}.

\section{Results}
\label{sec:results}

Table~\ref{tab:summary} provides an overview of the results of each hypothesis test; the subsections below report each in detail.

\begin{table*}[t]
\caption{Summary of pre-registered hypotheses and key findings. The exploratory row reports an unplanned observation. JP = Japanese participants; AM = American participants; HC = high-context; LC = low-context.}
\label{tab:summary}
\small
\begin{tabular}{p{1.8cm} p{3.6cm} p{6.0cm} p{0.6cm} p{2.3cm}}
\toprule
 & \textbf{Hypothesis} & \textbf{Key Result} & \textbf{Fig.} & \textbf{Verdict} \\
\midrule
\textbf{H0} & JP write more high-context emails than AM at baseline & JP wrote in a significantly more HC style than AM ($p < .001$) & \ref{fig:h0_baseline} & Supported \\
\addlinespace
\textbf{H1} & AI drafts shift the communication style of emails toward the draft's orientation & Emails shifted toward the HC style under the HC draft, and toward the LC style under the LC draft (both $p < .001$) & \ref{fig:h1_change} & Supported \\
\addlinespace
\textbf{H2} & Misaligned drafts produce larger shifts than aligned ones & Misaligned shift was 10--16$\times$ larger than aligned shift in both cultural groups (both $p < .001$) & \ref{fig:h2_change} & Supported \\
\addlinespace
\textbf{H3} & Aligned drafts produce higher AI reliance than misaligned ones & JP retained significantly more of the aligned draft ($p < .001$); AM showed no significant difference (n.s.) & \ref{fig:ai_reliance_context} & Partially supported \\
\addlinespace
\midrule
\textbf{Exploratory} & --- & JP generated AI drafts in both conditions at a substantially higher rate than AM ($96.6\%$ vs.\ $64.0\%$) & \ref{fig:draft_generated} & --- \\
\bottomrule
\end{tabular}
\end{table*}

\begin{figure}[h]
  \centering
  \includegraphics[width=0.6\linewidth]{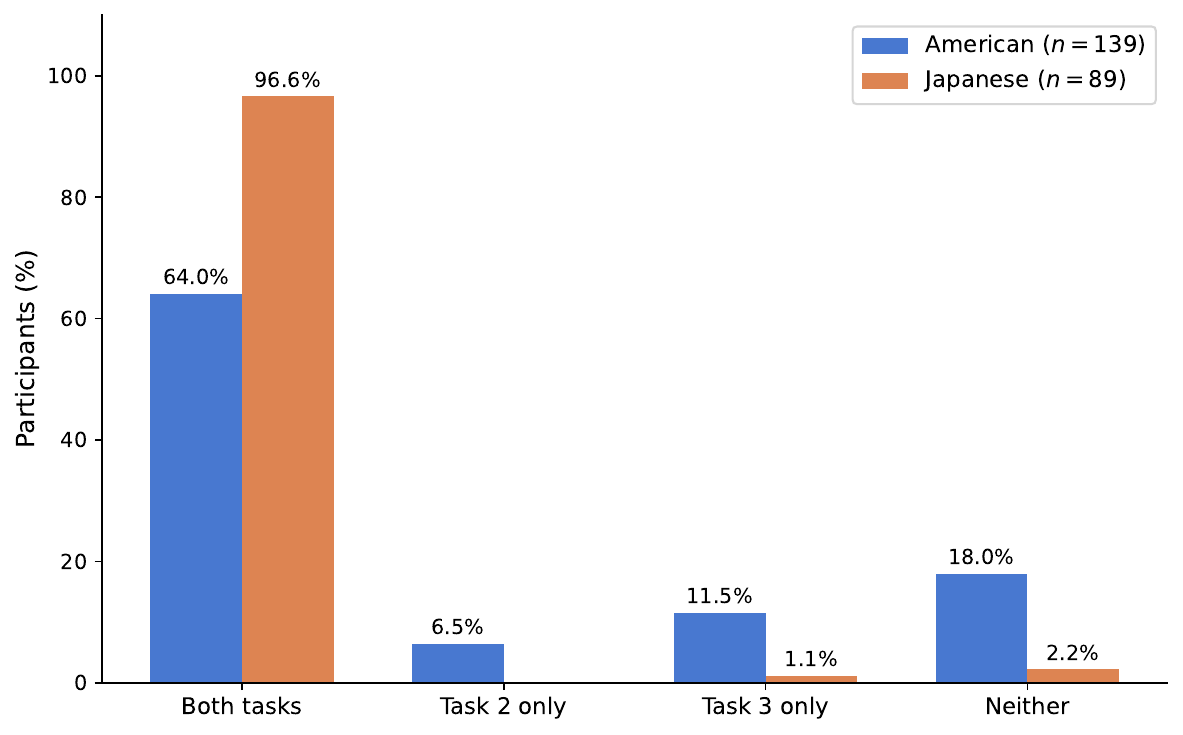}
  \caption{AI draft generation by task and cultural group. Task 1 served as a no-AI baseline and did not offer the Generate Draft option. Results shown are for Tasks 2 and 3 only.}
  \label{fig:draft_generated}
\end{figure}

Overall, 175 of 228 participants (76.8\%) clicked the ``Generate Draft'' button in both tasks (see Figure~\ref{fig:draft_generated}). We include those participants in the later analysis for H0 to H3 to maintain a consistent analytic sample across all hypotheses. A further 26 participants generated the draft in only one task (Task~2 only: $n = 9$, 3.9\%; Task~3 only: $n = 17$, 7.5\%), and 27 (11.8\%) did not generate a draft in either task. Interestingly, draft generation rates varied substantially across cultural groups: 86 of 89 Japanese participants (96.6\%) generated the draft in both tasks, compared to 89 of 139 American participants (64.0\%).

To understand this gap, we further examined open-ended responses from American participants who did not generate the AI draft in either task. These responses suggest that non-adoption by Americans was driven primarily by a preference for authentic self-expression and skepticism toward AI writing quality. For example, one participant noted \emph{``I felt like I wanted to write from how I really felt \ldots it feels more genuine,''} while another stated \emph{``I don't trust AI in cases that touch on very intricate matters.''} 

\subsection{H0: Baseline Differences Between Japanese and American Emails}

\begin{figure}[h]
  \centering
  \includegraphics[width=0.4\linewidth]{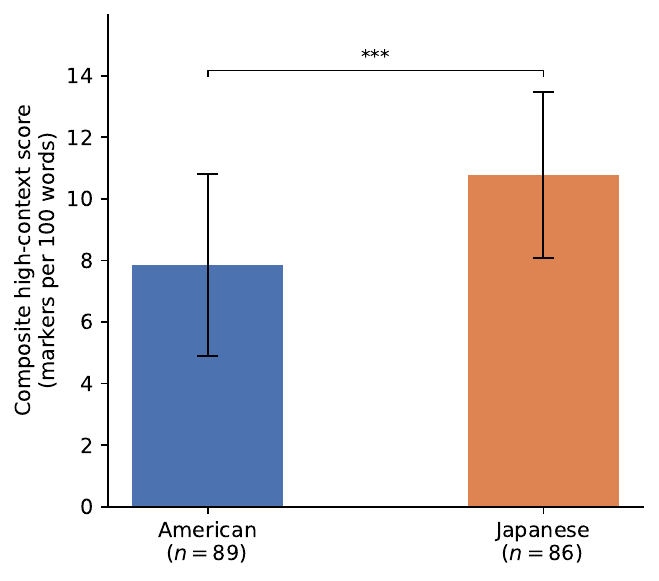}
  \caption{H0: Mean composite high-context scores (markers per 100 words) for American ($n=89$) and Japanese ($n=86$) participants in the no-AI baseline condition. Error bars show $\pm 1$ SD. $***$ $p<.001$ (independent-samples $t$-test, one-tailed).}
  \label{fig:h0_baseline}
\end{figure}

To test H0, we compared composite high-context scores (per 100 words) between American and Japanese participants in the no-AI baseline condition (Task~1).
Figure~\ref{fig:h0_baseline} shows the results ($n_{\text{AM}} = 89$, $n_{\text{JP}} = 86$). A one-tailed independent-samples $t$-test revealed that Japanese participants scored significantly higher on the composite high-context measure than American participants ($p < .001$). Exploratory analysis of individual marker categories suggests this difference was driven primarily by markers such as intensifiers (e.g., \textit{sincerely}, \textit{truly}), polite request forms (e.g., \textit{would it be possible to}), and apologies with a face-threatening act (e.g., \textit{I apologize for the inconvenience}). These results \textbf{support} H0: Japanese participants wrote in a more high-context style than American participants in the no-AI baseline.

\subsection{H1: Effect of AI Draft on Communication Style}

To test H1, we computed a change score for each AI-assisted email: the participant's total high-context marker score (per 100 words) in the AI task (Task~2 or~3) minus their score in the no-AI baseline (Task~1).
This yielded two change scores per participant, one for the low-context condition and one for the high-context condition.

As preregistered, we modeled change scores using a linear mixed-effects model (LMM) with a random intercept per participant to account for the fact that each participant contributed scores in both conditions.
The following fixed effects are included: AI draft condition (low-context vs.\ high-context), cultural group (American vs.\ Japanese), participants' baseline high-context score (original communication style), scenario, age, gender, AI use frequency, and AI reliance score.
Continuous predictors were mean-centered prior to fitting. We then tested whether each condition's model-estimated mean change score differed from zero using Wald $z$-tests derived from the fitted model.

\begin{figure}[h]
  \centering
  \includegraphics[width=0.4\linewidth]{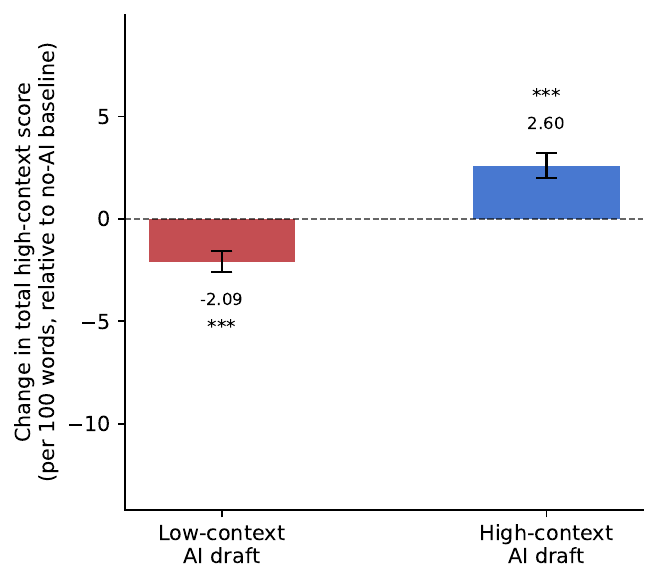}
  \caption{H1: Descriptive mean change in total high-context marker score (per 100 words) relative to the no-AI baseline, by AI draft condition ($n = 175$). Error bars show 95\% CIs. Significance vs.\ zero: $***$ $p < .001$ (Wald $z$-test from linear mixed-effects model).}
  \label{fig:h1_change}
\end{figure}

The results showed a clear directional pattern (Figure~\ref{fig:h1_change}).
Under the high-context AI draft, participants used significantly more high-context markers relative to their baseline ($M = +2.60$; $z = 6.47$, $p < .001$).
Under the low-context AI draft, participants used significantly fewer high-context markers ($M = -2.09$; $z = -6.22$, $p < .001$).

The LMM confirmed a strong effect of AI draft condition ($\beta = -4.75$, $SE = 0.19$, $z = -25.71$, $p < .001$): the low-context condition produced a 4.75-point lower change score than the high-context condition after adjusting for all covariates (descriptive means are shown in Figure~\ref{fig:h1_change}, not adjusted means).
Cultural group was also a significant predictor ($\beta_{\text{JP}} = -1.01$, $SE = 0.25$, $z = -4.08$, $p < .001$), indicating that Japanese participants showed a smaller overall shift relative to American participants across both conditions.
Baseline high-context score was a strong negative predictor ($\beta = -0.93$, $SE = 0.04$, $z = -25.65$, $p < .001$), indicating that participants with higher baseline scores showed smaller subsequent shifts.
Scenario had a significant effect for some levels, with participants showing larger shifts under the refusal and disagreement scenarios than under the favor-request scenario; age, gender, and AI use frequency were not significant ($p > .05$ for all).

These findings \textbf{support} H1: exposure to an AI email draft shifted the communication style expressed in participants' emails in the direction of the draft's context orientation.

\subsection{H2: Cultural Asymmetry in AI Influence}

To test H2, we extended the H1 model by adding a culture~$\times$~condition interaction term, while retaining all other fixed effects (original score, scenario, age, gender, AI use frequency, AI reliance) and the random intercept per participant ($n_{\text{AM}} = 89$, $n_{\text{JP}} = 86$).

The culture~$\times$~condition interaction was significant ($\beta = 1.50$, $SE = 0.35$, $z = 4.26$, $p < .001$), confirming that the effect of AI draft style on communication change differed between American and Japanese participants.
We then examined simple effects within each cultural group by fitting the same LMM separately to American and Japanese participants.

\begin{figure}[h]
  \centering
  \includegraphics[width=0.5\linewidth]{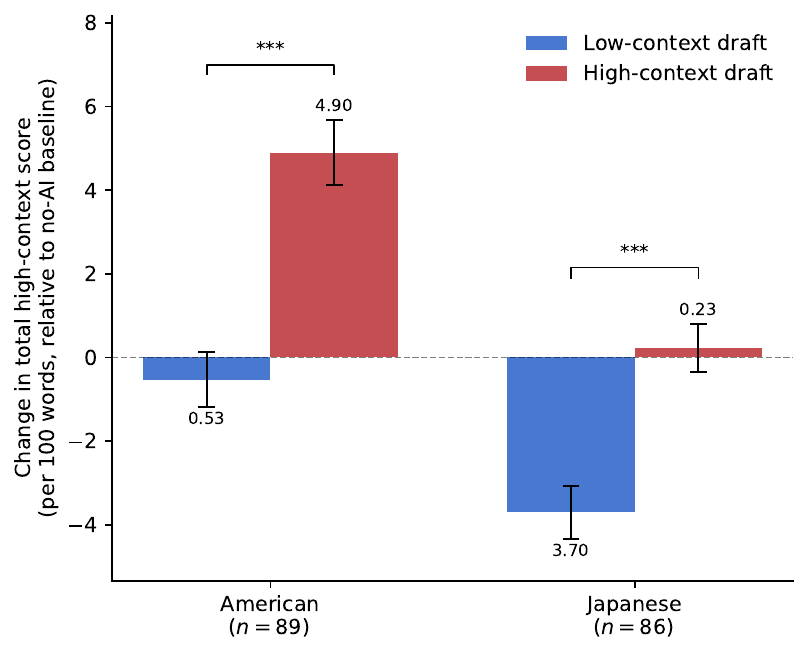}
  \caption{H2: Mean change in total high-context marker score (per 100 words) relative to the no-AI baseline, by cultural group and draft condition. Error bars show 95\% CIs. Significance brackets show within-group simple effects (low-context vs.\ high-context draft): $***$ $p < .001$ (per-culture LMM, one-tailed Wald $z$-test).}
  \label{fig:h2_change}
\end{figure}

Within Japanese participants, the misaligned (low-context) draft produced a substantially larger shift than the aligned (high-context) draft ($|M_{\text{LC}}| = 3.70$, $|M_{\text{HC}}| = 0.23$; $\beta = -4.03$, $SE = 0.24$, $z = -17.01$, $p < .001$, one-tailed).
Within American participants, the misaligned (high-context) draft also produced a substantially larger shift than the aligned (low-context) draft ($|M_{\text{HC}}| = 4.90$, $|M_{\text{LC}}| = 0.53$; $\beta = -5.44$, $SE = 0.26$, $z = -21.17$, $p < .001$, one-tailed).
In both cases, the shift was larger when the AI draft was culturally misaligned with the participant's background (Figure~\ref{fig:h2_change}).
Table~\ref{tab:email_examples} illustrates this pattern with representative emails showing a pronounced contrast between the no-AI and misaligned AI draft conditions for each cultural group.

These results \textbf{support} H2.
Both cultural groups shifted more under the misaligned draft than the aligned one: Japanese participants showed a larger shift under the low-context draft, and American participants showed a larger shift under the high-context draft.

\begin{table*}[t]
\small
\caption{Representative email excerpts from the disagreement scenario illustrating shifts in communication style across cultural groups and draft conditions. Japanese emails were originally written in Japanese and translated to English for analysis. \textcolor{teal}{Teal} highlights indirectness markers; \textcolor{orange}{orange} highlights Imposition Mitigation markers. Full emails for all four examples are provided in the supplemental material.}
\label{tab:email_examples}
\begin{tabular}{p{2.2cm}p{6.0cm}p{6.0cm}}
\toprule
 & \textbf{No AI} & \textbf{Misaligned AI Draft} \\
\midrule
\textbf{Japanese participants} &
``\textit{\textcolor{teal}{I believe} that a change of this nature could reduce the consistency of the dataset\ldots \textcolor{teal}{there is a concern that} the client \textcolor{teal}{may question} the validity of the reported results. \textcolor{teal}{I wanted to bring this to your attention promptly.} \textcolor{orange}{Thank you for your attention to this matter.}}''

\smallskip\noindent{\footnotesize (Participant: \texttt{12f76449})} &
``\textit{Changing the data collection method at this point will create inconsistencies among the data\ldots I propose either continuing with the current method through to the end, or testing the new method on a small scale in parallel.}''

\smallskip\noindent{\footnotesize (Low-context draft, Participant: \texttt{b4f2bcaa})} \\
\addlinespace
\textbf{American participants} &
``\textit{\textcolor{teal}{I think} we need to stick to our basics and revert changes. We need to ensure that we function properly and get the work completed.}''

\smallskip\noindent{\footnotesize (Participant: \texttt{2a9f7eb7})} &
``\textit{\textcolor{orange}{Thank you for your hard work.} \textcolor{teal}{I am writing to share a concern}\ldots \textcolor{orange}{I apologize for the inconvenience during your busy schedule,} but \textcolor{teal}{would it be possible to reconsider the timing and procedure of the change?} \textcolor{orange}{Thank you very much.}}''

\smallskip\noindent{\footnotesize (High-context draft, Participant: \texttt{88fa5dfc})} \\
\bottomrule
\end{tabular}
\end{table*}

\subsection{H3: AI Reliance and Cultural Alignment}

To test H3, we compared AI reliance scores within each cultural group across the two draft conditions using a one-tailed Wilcoxon signed-rank test.

\begin{figure}[h]
  \centering
  \includegraphics[width=0.6\linewidth]{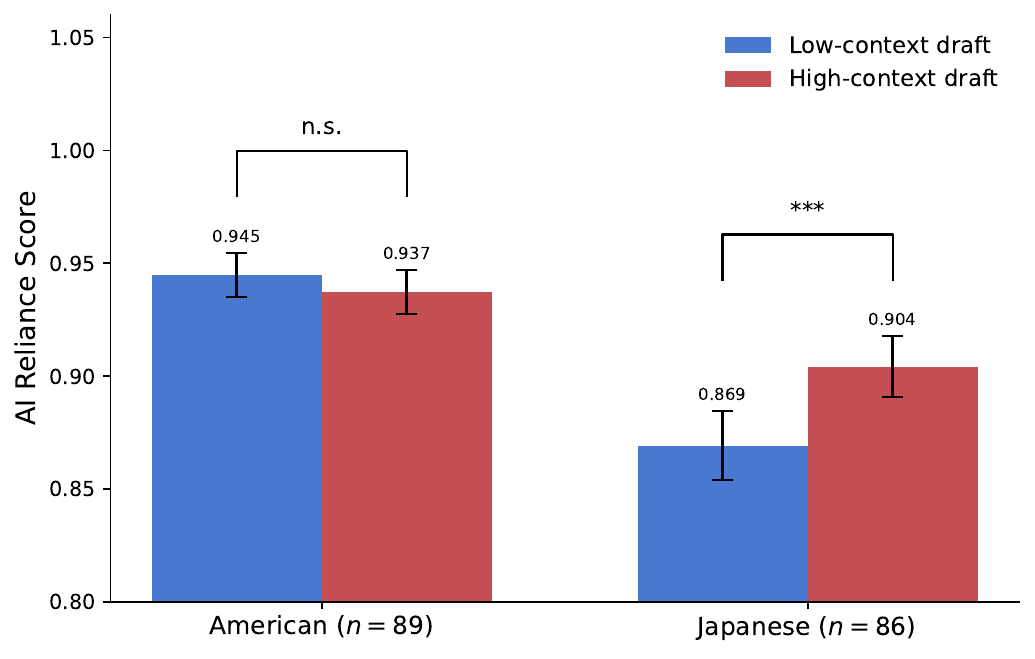}
  \caption{H3: Mean AI reliance scores by cultural group and draft condition. Error bars show $\pm 1$ SE. Significance (within-group, low vs.\ high-context): $***$ $p < .001$, n.s.\ not significant (Wilcoxon signed-rank test, one-tailed).}
  \label{fig:ai_reliance_context}
\end{figure}

Across both cultural groups and conditions, AI reliance scores were generally high (all means above 0.85), indicating that participants retained the majority of the AI draft text regardless of its cultural orientation (Figure~\ref{fig:ai_reliance_context}).
For Japanese participants, AI reliance was significantly higher in the high-context (culturally aligned) condition than in the low-context (misaligned) condition ($M_{\text{high}} = 0.904$, $M_{\text{low}} = 0.869$; Wilcoxon $T^{+} = 2322$, $p < .001$).
For American participants, reliance was numerically higher in the low-context (aligned) condition than in the high-context (misaligned) condition ($M_{\text{low}} = 0.945$, $M_{\text{high}} = 0.937$), but this difference was not significant ($T^{+} = 1058$, $p = .689$).

These results \textbf{partially support} H3.
Japanese participants retained significantly more of a culturally aligned draft, consistent with the prediction that alignment increases reliance.
American participants showed the expected direction but not a significant effect, which may in part reflect the generally high reliance scores in both conditions (both above 0.93), leaving limited room to detect a difference.

\section{Discussion}


We examined whether AI email drafts shift the cultural communication style of professional emails and whether this effect depends on the cultural alignment between the draft and the writer's cultural background.

The results of our experiment showed that \textbf{culturally distinct  communication norms in emails still exist}: Japanese emails were significantly more high-context, using more politeness markers and face-saving apologies, than American participants' emails. This finding is especially noteworthy because workplace globalization and the influence of technology is often feared to homogenize cultures~\cite{kwet2019digital,greig2002end,agarwal2025chi}, which our results suggest has not (yet?) happened. It is also noteworthy because decades of cross-cultural communication research have posited differences across low and high communication styles~\cite{hall1976beyond, brown1987politeness}, but direct empirical evidence of this difference in email communication remains limited. As such, our email dataset and analysis of communication style markers advance prior work in both anthropology and HCI, showing that workplace emails (favor-requests, disagreements, and refusals) are written significantly differently in Japan versus the US.  

However, in line with our hypotheses, we found that \textbf{using AI email drafts could shift culturally distinctive styles of communication in emails}: In our experiment, AI email drafts shifted users' written communication style away from their own cultural norms. Simply viewing and editing an AI-generated draft written in a particular communication style changed people's own writing in that direction, relative to what they produced without AI assistance. That the effect reversed with draft type confirms the communication style of the draft, not a general task effect, drove the changes. Importantly, the effect was consistent across both cultural groups and both draft types and especially strong when the AI draft was culturally misaligned with the participant's background. These results extend the AI writing homogenization literature from lexical and stylistic convergence ~\cite{agarwal2025chi} to cultural communication style in emails. 




Our experiment also confirmed our hypothesis that \textbf{culturally misaligned AI drafts produce larger shifts within each group} than culturally aligned AI drafts. Participants' emails did not simply shift towards whatever AI draft style they saw; they drifted \emph{more} when the draft clashed with their own cultural communication style. Importantly, this effect ran in both directions (the low-context draft shifted Japanese participants, while the high-context draft equally shifted American participants),  suggesting that AI's influence on email communication norms is not unique to Western-centric models but reflects a more general mechanism whereby AI can shape users' writing whenever its outputs differ from the cultural communication style of their background. While prior work has often focused on a Western/non-Western comparison (e.g., comparing Indian to US participants~\cite{agarwal2025chi,gao2026framing}), our findings show that the AI influence applies to both Americans and Japanese \emph{writing in their own native languages}---a first comparison of AI influence on two cultural groups from highly industrialized countries. This suggests that AI writing tools' homogenizing influence may generalize beyond populations shaped by colonial power relations, extending to cultural contexts like the US and Japan. 

Much of these effects are due to the human tendency to rely on AI suggestions. We found that retention of AI draft text was high across the board (AI reliance score $> 0.85$), but \textbf{Japanese participants leaned on the AI draft significantly more when it was culturally aligned (high-context) than when misaligned}. 
More precisely, Japanese participants retained significantly more of the aligned high-context draft (AI reliance score = 0.90) than the misaligned low-context one (0.87), while American participants showed no significant difference (aligned: 0.95, misaligned: 0.94). 
This suggests that Japanese participants may have been more willing to accept the AI draft when it ``sounded like them,'' in line with the asymmetric social cost of misalignment between the two cultures: in Japanese professional contexts, hierarchies are generally seen as more pronounced than in the US~\cite{kamins1998multi}. Submitting an insufficiently mitigating email to a superior risks genuine face-threat~\cite{gagne2010reexamining, lee2012cultural}, which may have compelled participants to edit more heavily; in American professional contexts, using more imposition mitigation than one's norm carries comparatively little social cost. If so, this asymmetric social cost may have directly reduced the magnitude of the Japanese shift.

Interestingly, \textbf{Japanese participants made use of the option to generate an AI email draft significantly more often than American participants (96.6\% vs.\ 64.0\%)}. A possible explanation lies in the results of a recent large-scale survey (Spring 2025, 25 countries), showing that Americans are considerably more concerned about AI than Japanese in general (50\% of Americans reported feeling more concerned than excited about AI's increasing use in daily life, compared to 28\% in Japan~\cite{pew2025ai}). This suggests that cultural differences in AI attitudes may shape not only how people use AI drafts, but whether they choose to engage with them at all.

The AI-induced drift in the cultural communication style expressed in participants' emails carries real stakes. First, cultural diversity in communication is itself a value. Different groups have developed distinct ways of communicating. A technology that systematically shifts users toward a single dominant style raises questions of representation: whose communication styles are preserved in AI-assisted emails, and whose are quietly displaced? 
Second, the effect we observed is unlikely to stop with the individual user who uses AI. Users whose communication style was overwritten by an AI email draft then send this email to others. Recipients may over time adapt their own expectations and norms in response to what they receive, known as mutual constitution of culture and psyche, in which cultural products such as emails shape inside-the-head communication norms over time~\cite{morling2016cultural}. Such shifts in culturally distinctive communication style in emails may therefore not remain confined to the individual writer: it may spread gradually across social networks, cumulatively reshaping how people across cultures write and communicate. According to Sourati et al.,  this concern extends beyond individual identity since LLM-driven standardization of language and reasoning threatens the epistemic benefits of human diversity more broadly~\cite{sourati2026homogenizing}.

\begin{figure*}[t]
\centering
\includegraphics[width=\textwidth]{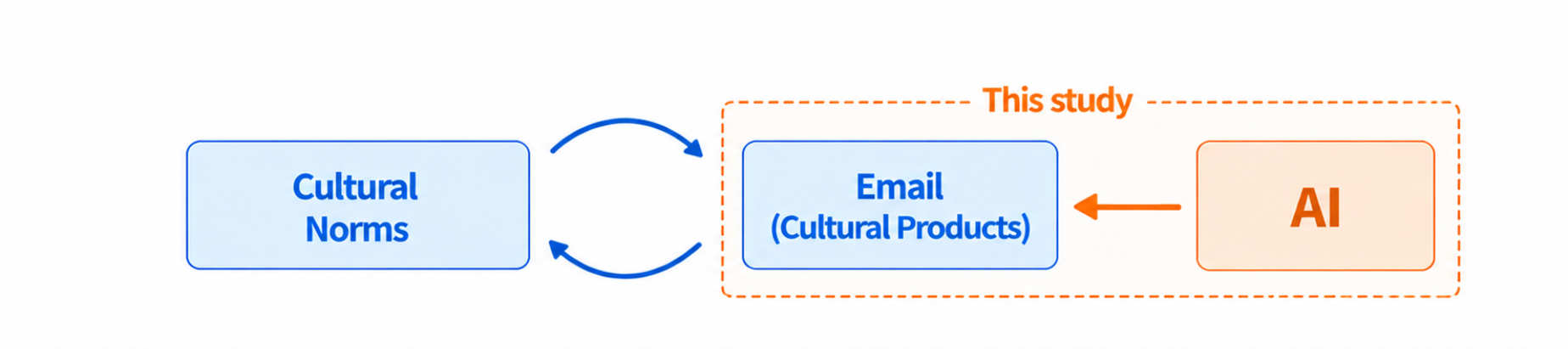}
\caption{The cultural cycle of communication norms, showing where AI email drafts intervene.}
\label{fig:cultural_cycle}
\end{figure*}

It is, of course, important to acknowledge that changes in cultural norms are a normal phenomenon: culture is a dynamic construct, continuously shaped by people's interactions with each other and with human-created artifacts~\cite{morling2016cultural}. As Figure~\ref{fig:cultural_cycle} illustrates, cultural norms shape how people write emails; and those emails, as cultural products, re-enter the social environment and contribute to norm evolution over time. AI email drafts intervene at the point of email writing. Our study did not directly measure changes in participants' underlying communication norms --- what we captured was a shift in the style of individual emails produced in a single session. But situated within this cultural cycle, that shift is meaningful: if AI drafts consistently push email style in a particular direction across many users and interactions, those emails accumulate as inputs into the cycle and may, over time, constitute a first step toward reshaping communication norms and values more broadly. With AI now integrated into everyday writing tools, it is prone to become a recurring force in this dynamic --- one that, unlike a typical human-to-human interaction, operates at scale and in a systematically biased direction. 

\subsection{Design Implications}
The fact that users retain the majority of AI-generated content regardless of cultural fit means that cultural biases embedded in AI systems can propagate quietly across populations without deliberate choice. This has direct implications for how AI writing tools should be designed, which we present in the following. 

\textbf{Adapting AI to cultural communication norms:} Our findings suggest that AI writing assistants should not treat communication style as a one-size-fits-all default, but as a dimension that can and should be personalized to a user's cultural background. We found that AI-drafted emails shift users' culturally distinctive communication styles, even when the draft is misaligned with their cultural background, which raises the risk of pulling communication norms towards the model's default. To counteract this, AI email drafts could infer or elicit user's cultural communication style (e.g., through onboarding, locale settings, or adaptive learning from prior edits) and calibrate draft directness, hedging, and politeness markers accordingly. Whether this adaptation should be automatic versus user-controlled remains an open question; an automatic adaptation can risk stereotyping, wrong inference~\cite{neplenbroek2025reading}, and even the perception of mimicry~\cite{basoah2025should}, while user control may be perceived as an additional burden to the user. Understanding these tradeoffs and how they might be perceived by different cultural groups will be an important step before adapting AI's cultural communication style.

\textbf{Rethinking what ``helpful'' AI writing assistance means:} Our asymmetric finding that AI drafts exert more influence when they are culturally misaligned with the user questions whether more fluent, more confident AI suggestions are unambiguously helpful. If misaligned drafts are the ones that shift users' style the most, the tools may be least trustworthy when they most need to be, since users have the least basis for detecting or resisting a mismatch they're not attuned to. This reframes the design goal: rather than optimizing solely for adoption or perceived helpfulness, designers may need to consider mechanisms that flag stylistic divergence to users, or add friction whenever there is high potential for a drift. 

\textbf{Signaling the AI's prevalent communication style:} In line with this, AI tools that support people's writing could make the cultural orientation of the tool visible to users, enabling them to engage more critically with AI-generated content rather than accepting it by default. Indeed, recent work has shown that a simple overview of the AI's biases can decrease user reliance on writing suggestions~\cite{gao2026framing}. Similar signifiers could be used to flag biases in communication norms. 

\textbf{Integrating cultural communication norms in evaluations of AI writing assistants:} Our work suggests that evaluation practices of AI writing assistants should incorporate cultural communication norms alongside standard measures like grammaticality, coherence, or task success. This is important for researchers and practitioners because benchmarks and user studies that only evaluate within a single cultural context risk missing the kind of asymmetric influence of AI bias that we observed in our study. Our email dataset and computational analyses for quantifying individual users' cultural communication styles can serve as a first step towards routinely assessing such AI biases. 


\section{Limitations and Future Work}

This study has several limitations.

First, translation was involved at multiple stages in the study, and this could introduce potential artifacts. This included the scenarios and AI email drafts, and Japanese participants' emails for scoring. Still, a Japanese--English bilingual speaker manually reviewed all translations, with particular attention to whether linguistic markers relevant to high- and low-context communication were preserved.

Second, our experiment is limited in its external validity. Our experiment was conducted online, and participants did not actually send the emails they produced. To increase ecological validity, we asked participants to imagine themselves genuinely in the sender's role, and a confirmation dialog before submission asked whether they would be willing to send the email to a real superior, ensuring the task carried some degree of felt accountability. Nonetheless, naturalistic email writing may differ from laboratory conditions in important ways: real emails involve ongoing relationships, actual professional stakes, and the freedom to choose whether to use AI assistance at all. Future work should examine AI draft adoption in situ to study how people engage with AI-generated drafts in their actual email clients.

Third, the high AI reliance observed across conditions, even under cultural misalignment, raises questions that this study cannot answer. It remains unclear whether this reflects cognitive ease (editing a draft is more demanding than accepting it), trust in AI writing quality, the efficiency afforded by having a ready-made text, or simply the experimental setup. Future work should examine the mechanisms driving AI draft acceptance, including whether users are aware of cultural mismatches and choose to accept the draft anyway, or whether misalignment goes unnoticed.

Additionally, the shifts observed in this study reflect communication style within a single experimental session. Whether repeated exposure to culturally misaligned AI drafts leads to durable changes in users' natural writing style, independent of AI assistance, remains unknown. Longitudinal designs that track writing style before, during, and after a period of AI writing tool use would be needed to address this question.

The present findings are also specific to workplace email writing across three scenario types --- all involving face-threatening acts in a hierarchical relationship. Whether similar effects emerge in other communication contexts (e.g., peer-to-peer emails, informal messages, or non-email genres) remains an open question for future work.

Finally, the study compared American and Japanese participants --- two of the most frequently studied pairings in cultural psychology research. Whether the patterns we observe generalize to other cultural groups at intermediate points on the high/low-context spectrum remains an open question.


\section{Conclusion}

This paper presents an online experiment with participants from Japan and the US---two groups that are known to have culturally distinct communication norms. Our findings demonstrate that AI email drafts quietly reshape the culturally distinctive communication style that users bring to professional writing, with the effect being most powerful when the AI and user are culturally misaligned. This points to a need for writing tools that are designed not as culturally neutral assistants, but as systems that acknowledge and adapt to the diversity of how people communicate.

\bibliographystyle{ACM-Reference-Format}
\bibliography{references}

\appendix
\section{Power Analysis}
\label{app:power}

Sample size was determined to detect the culture $\times$ AI
condition interaction, which is harder to detect than a main effect alone. We assumed a medium effect size (Cohen's
$f = 0.25$), two-tailed $\alpha = .05$, and target power of $.80$.
To account for the repeated-measures structure (three within-subjects
conditions, assumed correlation $\rho = .50$ among repeated measures),
the effect size was adjusted following standard mixed-design formulas,
yielding an adjusted $f \approx 0.31$. A power analysis conducted in
Python using \texttt{statsmodels} \texttt{FTestAnovaPower} indicated
that 86 participants per group are required to achieve power $= .801$.

\section{Participant Demographics}
\label{app:demographics}

Please refer to Table~\ref{tab:demographics} for detailed demographic information by cultural group.

\begin{table*}[h]
\caption{Participant demographics by cultural group. Percentages are out of the group total.}
\label{tab:demographics}
\small
\begin{tabular}{lcc}
\toprule
& \textbf{American} ($n = 89$) & \textbf{Japanese} ($n = 86$) \\
\midrule
\textbf{Age} & & \\
\quad $M$ (SD) & 38.0 (11.0) & 41.3 (10.6) \\
\quad Range & 20–79 & 24–63 \\
\addlinespace
\textbf{Gender} & & \\
\quad Male & 41 (46.1\%) & 55 (64.0\%) \\
\quad Female & 47 (52.8\%) & 30 (34.8\%) \\
\quad Non-binary & 1 (1.1\%) & 1 (1.2\%) \\
\addlinespace
\textbf{Highest education} & & \\
\quad Some high school (no diploma) & 1 (1.1\%) & 0 \\
\quad High school diploma & 13 (14.6\%) & 9 (10.5\%) \\
\quad Some university (no degree) & 15 (16.9\%) & 3 (3.5\%) \\
\quad Bachelor's degree & 30 (33.7\%) & 68 (79.0\%) \\
\quad Some graduate school (no degree) & 2 (2.2\%) & 0 \\
\quad Graduate degree (Master's / PhD) & 28 (31.5\%) & 6 (7.0\%) \\
\addlinespace
\textbf{Tenure at current company} & & \\
\quad Less than 1 year & 4 (4.5\%) & 9 (10.5\%) \\
\quad 1–3 years & 18 (20.2\%) & 17 (19.8\%) \\
\quad 3–5 years & 20 (22.5\%) & 18 (20.9\%) \\
\quad 5–10 years & 20 (22.5\%) & 16 (18.6\%) \\
\quad More than 10 years & 27 (30.3\%) & 26 (30.2\%) \\
\addlinespace
\textbf{AI writing tool use} & & \\
\quad Never & 14 (15.7\%) & 10 (11.6\%) \\
\quad Rarely & 16 (18.0\%) & 1 (1.2\%) \\
\quad Sometimes & 28 (31.5\%) & 29 (33.7\%) \\
\quad Often & 18 (20.2\%) & 12 (14.0\%) \\
\quad Very often & 13 (14.6\%) & 34 (39.5\%) \\
\bottomrule
\end{tabular}
\end{table*}

\section{Study Design}
\label{app:design}

Please refer to Table~\ref{tab:design} for the full Latin square counterbalancing of scenario type and AI draft order.

\begin{table*}[h]
\caption{Study design: Latin square counterbalancing of scenario type and AI draft order.}
\label{tab:design}
\small
\begin{tabular}{lccc}
\toprule
& \textbf{Task 1 (No-AI)} & \textbf{Task 2 (AI)} & \textbf{Task 3 (AI)} \\
\midrule
Group A, Condition A & Favor-request & Disagreement + Low  & Refusal + High \\
Group A, Condition B & Favor-request & Disagreement + High & Refusal + Low  \\
Group B, Condition A & Disagreement  & Refusal + Low       & Favor-request + High \\
Group B, Condition B & Disagreement  & Refusal + High      & Favor-request + Low  \\
Group C, Condition A & Refusal       & Favor-request + Low & Disagreement + High \\
Group C, Condition B & Refusal       & Favor-request + High & Disagreement + Low \\
\bottomrule
\end{tabular}
\end{table*}

\section{Scenario and Draft Selection Procedure}
\label{app:scenarios}

The goal of this procedure was to identify, for each of the three scenario types (favor-request, disagreement, and refusal), the candidate scenario that would yield the largest style contrast between its high- and low-context AI drafts. This maximizes the strength of the manipulation, giving it the best opportunity to produce an observable effect on participants' writing. Drafts were generated by LLMs rather than crafted manually by the authors to ensure higher ecological validity.

For each of the three scenario types, three candidate scenarios were generated using Claude Sonnet 4.5 and manually verified to confirm that each scenario was a clear example of the target type. This yielded nine candidate scenarios in total (three per type).

For each of the nine candidate scenarios, six AI email drafts were generated: one per combination of three LLMs (GPT-5.5, Claude Sonnet 4.6, and Gemini 2.5 Flash, the most widely used generative AI tools not only in the United States but also in Japan~\cite{gmo2025genai}) and two prompt languages (English and Japanese). To determine which candidate scenario yielded the largest style contrast, each draft was scored on the coding scheme described in Section~\ref{sec:measures}. Japanese-prompt email drafts were translated into English prior to scoring so that all drafts were evaluated using a single, consistent coding scheme to preserve construct validity. For translation, we used GPT-5.5 Instant, and back-translated the resulting English drafts into Japanese using Claude Sonnet 5. Different models were used for each direction to ensure translational independence, preventing the back-translation model from reproducing the source text rather than translating independently. A Japanese--English bilingual speaker then manually reviewed all translations, with particular attention to whether high-context cues and low-context cues were faithfully preserved. Each draft was then rated using the composite high-context scoring scheme described in Section~\ref{sec:measures}, yielding a total high-context score per draft. For each candidate scenario, the contrast was defined as the difference between the highest-scoring and lowest-scoring draft among the six.

For each scenario type, the candidate scenario with the largest contrast score was selected as the study stimulus. The highest-scoring draft for that scenario was assigned as the high-context AI draft, and the lowest-scoring draft was assigned as the low-context AI draft. This procedure ensures that the stimuli used in the experiment represent cases where the low-context and high-context drafts are maximally distinguishable, giving the manipulation the best opportunity to produce an observable effect on participants' writing. Table~\ref{tab:scenarios} presents the three selected scenarios used in the main study.

\begin{table*}[h]
\caption{The three workplace scenarios used in the study, one per scenario type.}
\label{tab:scenarios}
\small
\begin{tabular}{p{2.2cm}p{13.3cm}}
\toprule
\textbf{Type} & \textbf{Scenario} \\
\midrule
Favor-request &
  A close friend is getting married this Saturday. You need to travel out of town from Thursday evening through Sunday — effectively three working days away. You only found out last week that you are needed as part of the wedding party, so the request is coming on very short notice. Your boss has a known policy of not approving last-minute leave requests, and similar requests from colleagues have been turned down in the past. Missing your close friend's wedding is not an option for you. \emph{Write an email to your boss requesting three days of leave this week.} \\
\addlinespace
Disagreement &
  In today's team meeting, your boss outlined a plan to switch to a different data collection method midway through the current project. Based on your experience with the previous phase, you believe this change will introduce inconsistencies in the dataset that will make the final analysis unreliable. If the project proceeds this way, the results may not hold up to scrutiny when presented to stakeholders. \emph{Write an email to your boss expressing your concern about this direction.} \\
\addlinespace
Refusal &
  Your boss has just asked you to lead a new client proposal that needs to be submitted by end of next week. You are currently managing two ongoing projects, both with critical deadlines within the next ten days. Taking on the proposal would require approximately 15 additional hours of work, meaning you would not be able to meet your existing deadlines without serious quality trade-offs. \emph{Write an email to your boss explaining that you are not able to take on the new project at this time.} \\
\bottomrule
\end{tabular}
\end{table*}

Table~\ref{tab:draft_scores} reports average high-context scores for the 54 generated drafts by model and prompt language. GPT-5.5 produced the most low-context output ($M = 7.33$ markers per 100 words), followed by Gemini 2.5 Flash ($M = 9.74$) and Claude Sonnet 4.6 ($M = 10.27$). Drafts generated from Japanese prompts were substantially more high-context than those generated from English prompts ($M = 10.48$ vs.\ $M = 7.75$), indicating that prompt language itself shapes the context-orientation of the output, independent of the model used.

\begin{table}[h]
\caption{Average high-context scores of the AI-generated drafts by model and prompt language (markers per 100 words).}
\label{tab:draft_scores}
\small
\begin{tabular}{lccc}
\toprule
 & \textbf{English prompt} & \textbf{Japanese prompt} & \textbf{Overall} \\
\midrule
GPT-5.5           & 6.57 & 8.08  & \textbf{7.33}  \\
Gemini 2.5 Flash  & 8.02 & 11.47 & \textbf{9.74}  \\
Claude Sonnet 4.6 & 8.65 & 11.89 & \textbf{10.27} \\
\midrule
\textbf{Overall}  & \textbf{7.75} & \textbf{10.48} & \\
\bottomrule
\end{tabular}
\end{table}

\section{AI Email Drafts}
\label{app:drafts}

The following presents the six English AI email drafts used in the study (one low-context and one high-context draft per scenario).

\subsection*{Favor-Request}

\textbf{Low-context draft} \hfill \textit{Gemini, English prompt}

\begin{quote}\small
\textit{Subject: Request for Leave This Week}

Hi [Supervisor's Name],

I hope this email finds you well. I am writing to request three days of leave from Wednesday, November 8th, through Friday, November 10th.

A very close friend is getting married this Saturday, and I was only recently asked to be a part of the wedding party. This requires me to travel out of town early to participate in necessary preparations.

I understand this is incredibly short notice and goes against your usual policy for last-minute requests, which I fully respect. However, missing this significant event for a close friend is truly not an option for me.

I am already proactively preparing for my absence by completing urgent tasks and ensuring comprehensive handovers for all critical responsibilities. I am committed to minimizing any disruption.

Thank you for considering my urgent request.

Sincerely, [Your Name]
\end{quote}

\textbf{High-context draft} \hfill \textit{Gemini, Japanese prompt}

\begin{quote}\small
\textit{Subject: Request for Leave This Week}

Hi [Supervisor's Name],

I apologize for contacting you while you are away. I sincerely apologize for this sudden message, but I am writing to request three days of leave from this Thursday evening through Sunday.

A long-time close friend's wedding is this Saturday, and last week I was suddenly asked to attend as a member of the wedding party. This has resulted in a very last-minute request, and I am deeply sorry.

I am fully aware that this last-minute leave request goes against your policy, and I am truly sorry. However, I humbly ask for your permission so that I may fulfill this important role in celebrating my close friend's new chapter.

To ensure there is no disruption to work during my absence, I will complete my projects as much as possible in advance, ask my colleague [Name] to handle urgent matters, and ensure a thorough handover.

I apologize for this selfish request, but I sincerely hope you will approve it.

Respectfully, [Your Name]
\end{quote}

\subsection*{Disagreement}

\textbf{Low-context draft} \hfill \textit{GPT-5.5, English prompt}

\begin{quote}\small
\textit{Subject: Concern About Switching Data Collection Methods}

Hi [Supervisor's Name],

Thanks for asking to be contacted by email while you're away. After today's meeting I wanted to flag a concern about switching data-collection methods mid-project. Based on our experience in the first phase, changing methods now will introduce inconsistencies between datasets and impede comparability, increasing the likelihood that our final analysis will be unreliable and may not withstand stakeholder scrutiny.

I recommend either maintaining the current method through completion or running a small parallel pilot of the new method while documenting differences for analysis adjustments. I can prepare a brief impact assessment and proposed options for your review. Would you like me to proceed?

Best regards, [Your Name]
\end{quote}

\textbf{High-context draft} \hfill \textit{Claude, Japanese prompt}

\begin{quote}\small
\textit{Subject: Concern About Switching Data Collection Methods}

Hi [Supervisor's Name],

Thank you for your hard work.

I am writing to share a concern regarding the change to the data collection method explained in today's meeting.

A change midway through the project may make it difficult to maintain consistency with existing data. As a result, I feel there is a risk that the reliability of the final analysis will decrease, and that the validity of the results may be questioned when reporting to the client.

I apologize for the inconvenience during your busy schedule, but would it be possible to reconsider the timing and procedure of the change?

Thank you very much.

[Your Name]
\end{quote}

\subsection*{Refusal}

\textbf{Low-context draft} \hfill \textit{GPT-5.5, English prompt}

\begin{quote}\small
\textit{Subject: Regarding the New Client Proposal}

Hi [Supervisor's Name],

Per your request to communicate by email while you're away: thank you for considering me to lead the new client proposal. I'm currently managing two projects with critical deadlines within the next ten days; taking on the proposal would require about 15 additional hours and would force substantial quality trade-offs or missed deliverables on those projects. Given that, I can't accept the proposal assignment at this time without risking our commitments.

I can, however, help by outlining the proposal framework, mentoring a colleague to lead it, or starting a first draft for handoff. Please let me know which alternative you prefer or if you'd like to revisit this after next week.

Best regards, [Your Name]
\end{quote}

\textbf{High-context draft} \hfill \textit{Claude, Japanese prompt}

\begin{quote}\small
\textit{Subject: Regarding the New Client Proposal}

Hi [Supervisor's Name],

Thank you for your hard work.

I have received your request regarding the new client proposal. I would very much like to help, but I currently have two projects on hand, both with important deadlines within the next ten days.

I estimate that approximately 15 additional hours will be needed for this proposal, and it would be difficult to maintain the quality and meet the deadlines of my existing work.

I sincerely apologize, but I find it difficult to take this on at this time. Would you consider asking someone else? I am happy to assist with anything I can.

Thank you very much.

[Your Name]
\end{quote}

\section{Linguistic Markers}
\label{app:markers}

Tables~\ref{tab:d1markers} and~\ref{tab:d2markers} list the linguistic markers for the two components of the composite high-context score --- indirectness and imposition mitigation --- with a concrete example of each marker and its direct equivalent without the marker.

\begin{table*}[h]
\caption{Indirectness markers}
\label{tab:d1markers}
\small
\begin{tabular}{p{2.2cm}p{4.3cm}p{4.5cm}p{4.5cm}}
\toprule
\textbf{Marker} & \textbf{Definition} & \textbf{Example (with marker)} & \textbf{Direct equivalent (without marker)} \\
\midrule
Hedging words &
  Vague or uncertain language that softens a statement, signalling the speaker is not fully committed to a claim. &
  \emph{``I'm concerned that the timeline may be difficult to meet.''} &
  \emph{``The timeline cannot be met.''} \\
\addlinespace
Modal hedges &
  Modal verbs or constructions that reduce the forcefulness of a claim, making it tentative rather than assertive. &
  \emph{``This might create some difficulties for the team.''} &
  \emph{``This will create difficulties for the team.''} \\
\addlinespace
Approximators &
  Words expressing approximation rather than precision, reducing commitment to an exact figure or timeframe. &
  \emph{``I estimate approximately 15 additional hours will be needed.''} &
  \emph{``I need 15 more hours.''} \\
\addlinespace
Indirect speech acts &
  Using one speech act to perform another — e.g., framing a refusal as a consultation, or expressing disagreement as gratitude or a question. &
  \emph{``I am writing to consult you regarding a concern.''} (= I disagree with this decision) &
  \emph{``I disagree with this decision.''} \\
\addlinespace
Implicit disagreement / unclear refusal &
  Communicating disagreement or refusal without ever explicitly stating ``I disagree'' or ``I refuse''; the reader must infer the intent. &
  \emph{``I wanted to flag this before proceeding.''} (= I think your figures are wrong) &
  \emph{``I believe your figures are incorrect — here is the correct data.''} \\
\addlinespace
Passive / impersonal constructions &
  Removing the human agent through passive voice or impersonal constructions so consequences are described without assigning blame. &
  \emph{``There is a risk of conveying incorrect information to senior management.''} &
  \emph{``You will give them incorrect information.''} \\
\addlinespace
Rhetorical questions &
  Questions that do not expect a direct answer but perform a speech act — disagreement, warning, or pushback — while maintaining an interrogative surface form. &
  \emph{``Is this really the direction we want to take?''} (= I think this is the wrong direction) &
  \emph{``I think this is the wrong direction.''} \\
\addlinespace
Understatement &
  Describing something as less significant or extreme than it actually is as a face-saving strategy. &
  \emph{``I have a capacity concern this week.''} (= I cannot do this) &
  \emph{``I cannot take this on.''} \\
\bottomrule
\end{tabular}
\end{table*}

\begin{table*}[h]
\caption{Imposition Mitigation markers}
\label{tab:d2markers}
\small
\begin{tabular}{p{2.2cm}p{4.3cm}p{4.5cm}p{4.5cm}}
\toprule
\textbf{Marker} & \textbf{Definition} & \textbf{Example (with marker)} & \textbf{Direct equivalent (without marker)} \\
\midrule
Apology with face-threatening act &
  Apologising when performing a face-threatening act --- such as making a request, declining a superior's offer, or disagreeing with a superior's opinion --- explicitly acknowledging the social cost of the imposition or challenge. &
  \emph{``I sincerely apologize for the inconvenience.''} &
   \\
\addlinespace
Softeners &
  Linguistic devices inserted around a request to reduce its directness without changing its grammatical form. &
  \emph{``If it is not too much trouble, could you review this draft?''} &
  \emph{``Please review this draft.''} \\
\addlinespace
Polite request forms &
  Grammatically polite forms for making requests — questions or conditional structures used instead of imperatives. &
  \emph{``Would it be possible to reconsider the timing of the change?''} &
  \emph{``Please reconsider the timing of the change.''} \\
\addlinespace
Appreciation / gratitude &
  Expressing thanks or appreciation, as a face-saving preface or closing that frames the interaction positively. &
  \emph{``Thank you for thinking of me for this opportunity — I truly appreciate it.''} &
   \\
\addlinespace
Boundary acknowledgment &
  Explicitly recognising the imposition or constraint the request places on the recipient, demonstrating awareness of their situation. &
  \emph{``I completely understand you have a demanding schedule right now with the client project.''} &
   \\
\addlinespace
Opt-out offer &
  Explicitly offering the recipient an easy way to decline, making refusal socially permissible. &
  \emph{``Please know there is absolutely no pressure — I completely understand if this is not possible.''} &
   \\
\addlinespace
Minimizers &
  Language that downplays the size or effort of what is being requested, making the imposition seem smaller. &
  \emph{``Would you be willing to write a brief letter of recommendation?''} &
  \emph{``Would you be willing to write a detailed recommendation letter describing my achievements?''} \\
\addlinespace
Temporal softeners &
  Expressions that give the recipient flexibility in timing, reducing urgency. &
  \emph{``Please let me know at your convenience.''} &
  \emph{``Please respond by end of day Friday.''} \\
\addlinespace
Deference to authority &
  Language that explicitly places decision-making authority with the recipient, signalling subordination. &
  \emph{``Please let me know how you'd like to proceed — I will follow your judgment.''} &
  \emph{``I will proceed with Option A unless I hear otherwise.''} \\
\addlinespace
Indebtedness &
  Expressing a sense of personal obligation or emotional cost, signalling that the sender feels the weight of the imposition personally. &
  \emph{``It pains me to ask at such a busy time, but I would be very grateful for your support.''} &
  \emph{``I would like to request your support for this application.''} \\
\addlinespace
Ritual phrases &
  Conventional opener or closer phrases serving a social and relational function. &
  \emph{``I hope this email finds you well.''} &
   \\
\addlinespace
Intensifiers &
  Intensifying adverbs (\emph{sincerely}, \emph{deeply}, \emph{truly}, etc.) appearing directly before apology and gratitude, giving extra weight beyond a routine expression. &
  \emph{``I \textbf{sincerely} apologize.''} / \emph{``I \textbf{deeply} appreciate your help.''} &
  \emph{``I apologize.''} / \emph{``I appreciate your help.''} \\
\bottomrule
\end{tabular}
\end{table*}

\section{Post-Task Survey Questions}
\label{app:survey}

The following questions were presented to all participants after the three
email tasks, in the order listed.

\paragraph{Draft preference.}
``Looking back at the two AI-generated drafts you received — one in Task~2
and one in Task~3 — which felt more natural or appropriate for the
situation?'' (select one)
\begin{itemize}
  \item The draft from Task~2 felt more natural
  \item The draft from Task~3 felt more natural
  \item Both felt equally natural
  \item Neither felt natural
  \item I did not use the AI draft in Task~2
  \item I did not use the AI draft in Task~3
  \item I did not use the AI draft in either task
\end{itemize}

\paragraph{AI use frequency.}
``How often do you use AI tools to write emails (e.g., ChatGPT, Claude,
Copilot)?'' (select one)
\begin{itemize}
  \item Never
  \item I've only used them once or twice to write emails
  \item I use them occasionally to write emails
  \item I use them at least once a week to write emails
  \item I use them at least once daily to write emails
\end{itemize}

\paragraph{Job tenure.}
``How long have you been working at your current company?'' (select one)
\begin{itemize}
  \item Less than 1 year
  \item 1–3 years
  \item 3–5 years
  \item 5–10 years
  \item More than 10 years
\end{itemize}

\paragraph{Demographics.}
Age (open numeric entry); gender (Female / Male / Non-binary / Prefer not
to disclose); highest level of formal education (9-point scale from
\emph{No formal education} to \emph{Complete graduate degree}).

\section{Draft Preference by Cultural Alignment}
\label{app:preference}

Table~\ref{tab:preference} shows draft preference responses for the 175 participants included in the main analysis (those who generated AI drafts in both tasks). Preferences are classified as \emph{aligned} (preferred the culturally congruent draft: low-context for Americans, high-context for Japanese), \emph{misaligned} (preferred the culturally incongruent draft), \emph{both} (equally natural), or \emph{neither} (neither felt natural).

\begin{table*}[h]
  \centering
  \caption{Draft preference by cultural group ($n = 175$). Aligned = preferred the culturally congruent draft; misaligned = preferred the culturally incongruent draft.}
  \label{tab:preference}
  \begin{tabular}{lcccc}
    \toprule
    & Aligned & Misaligned & Both & Neither \\
    \midrule
    American ($n = 89$) & 27 (30\%) & 20 (22\%) & 36 (40\%) & 4 (5\%) \\
    Japanese ($n = 86$) & 42 (49\%) & 16 (19\%) & 20 (23\%) & 8 (9\%) \\
    \bottomrule
  \end{tabular}
\end{table*}

\end{document}